\documentclass[twocolumn,prl,float]{revtex4}

\usepackage{graphicx} 
\usepackage{float}
\usepackage{mhchem} 
\usepackage{siunitx}
\usepackage{lipsum} 
\usepackage{epsfig}
\usepackage{ragged2e}
\usepackage{hyperref}
\usepackage{amsmath,amsfonts,amsthm,bm}
\usepackage{xcolor}

\begin{document}



\author{Lorenzo Agosta}

\affiliation{Department of Chemistry, $\AA$ngstr\"{o}m Laboratory, Uppsala University, 751 21 Uppsala, Sweden}

\email{lorenzo.agosta@kemi.uu.se}



\author{Kersti Hermansson}

\affiliation{Department of Chemistry, $\AA$ngstr\"{o}m Laboratory, Uppsala University, 751 21 Uppsala, Sweden}

\author{Mikhail Dzugutov}

\affiliation{Department of Chemistry, $\AA$ngstr\"{o}m Laboratory, Uppsala University, 751 21 Uppsala, Sweden}





\title{Water under hydrophobic confinement: entropy and diffusion.}


\begin{abstract}
  The properties of liquid water are known to change drastically in confined geometries. A most interesting and intriguing phenomenon is that the diffusion of water is found to be strongly enhanced by the proximity of a hydrophobic confining wall relative to the bulk diffusion.  We report a molecular dynamics simulation using a classical water model investigating the water diffusion near a non-interacting smooth confining wall, which is assumed to imitate a hydrophobic surface. A pronounced diffusion enhancement is observed in the water layers adjacent to the wall. We present evidence  that the observed diffusion enhancement can be accounted for, with numerical accuracy, using the universal scaling law for liquid diffusion that relates the liquid diffusion rate to the excess entropy.  These results show that the scaling law, that has so far only been used for the description of the diffusion in simple liquids, can successfully describe the diffusion in water. It is thus shown that the law can be used for the description of  water dynamics under nano-scale confinement which is currently a subject of intense research activity.
\end{abstract}
\maketitle

Water, in its liquid form,  is the most ubiquitous liquid on our planet, but many of its features remain poorly understood and unexplored despite the immense ammount of research activity investigating different properties of this liquid \cite{structure_anomalies_water, rev_water_general, water_revised}. In addition to the technological and practical implications\cite{banuelos_oxide_2023,barry_advanced_2021}, these studies address a number of unique properties of water which are of great interest for the statistical physics of condensed matter . One aspect of liquid water that is currently being the focus of research activity is its behavior in confined geometries.
Water confined by geometrical constraints exhibits different structural features and dynamical behavior than in the bulk state \cite{cerveny16,nanofluids_interfaces}. The effect of confinement of water is largely investigated because of its relevance in several technological applications, such as ionic transport \cite{transport_water_nanotube}, ice nucleation control \cite{ice_nuclation}, friction \cite{friction}, and hydrophobicity\cite{water_dyn_hydrophobic,hydrophobic_nanopore, condesation_nanopore, dewetting}. Quasi-two-dimensional confined water have been actively studied in computer simulations\cite{confined_simulations} and real experiments\cite{corti_structure_2021,water_slip_nanotubes}. For instance, it has been observed that water in carbon nanotubes\cite{nanotube_rev} or silica nano-slits \cite{diff_polarity} exhibits dynamic anomalies. A peculiar observation is that in carbon nanotubes the molecular water diffusion undergoes a transition to be enhanced with respect to the bulk reference values (for diameters between 2 and 3 nanometers) to be strongly suppressed  diameters below 2 nanometers \cite{diff_nanotube}. This transition is currently object of investigation because it can be taken as model for understanding the influence of atomic structure on molecular diffusivity \cite{water_slip_nanotubes} and ionic transport in cell membranes (ref). Enhanced diffusion dynamics was also observed experimentally and from means of Molecular Dynamics (MD) simulations for water between graphene sheets \cite{carbone21, cicero}, clay materials\cite{gonzales08}, hydrophobic coatings \cite{water_pattern_diff}, water-vapor interface\cite{diff_water_vap}, and ideal hydrophobic surfaces \cite{rosky94, scheidler,stanley,agosta_entropic_2024}. The opposite behavior, when diffusion is reduced with respect to the bulk, is observed for hydrophilic surfaces \cite{drossel} as for the case of graphene-oxidized surfaces \cite{graphene_ox} and metal oxide interfaces\cite{water_silica_diff, diffusion_interface_rev, Agosta2017,debenedetti09,debenedetti09-2,supercooled}, with the former being triggered by exposing local surface dipole groups and the latter being promoted by the under-coordinated surface metal atoms. Another intriguing scenario is represented by the hydrophobicity induced by ordered water layers absorbed on hydrophilic surfaces\cite{Zhang21}. It was demonstrated that special patterns of water can self-assemble in a dense and locked monolayer structure that impedes the interaction with the surrounding liquid water \cite{Qi19,Wang19,chandler13,agosta_ceo2water}. In all theses cases, a deep understating of the anomalous water dynamics generated at the confining interface is missing and it would be of great interest in order to tailor and control this mechanism. 

Recently we have shown \cite{agosta_entropic_2024} that the enhanced diffusion experienced by a simple Lennard Jones liquid close to a non-interacting flat surface is quantitatively accounted by using an earlier reported scaling law for liquid diffusion that relates the diffusion rate to the local excess entropy \cite{dzugutov1,dzugutov2}. This law, however, has never been shown to describe water (or molecular liquid) diffusion, but only simple isotropic liquids.  
In the present study we adopt molecular dynamics (MD) simulations of a classical model of water confined by a smooth purely repulsive wall which is assumed to imitate a hydrophobic surface. A significant enhancement of the lateral diffusion in the layers adjacent to the wall is observed.  We demonstrate that the this effect can be quantitatively accounted for using the entropic mechanism as follows from the scaling law for liquid diffusion\cite{dzugutov1,dzugutov2}. These results also imply that the application scope of the scaling law, that  was successfully tested  on water in this study, can be extended to the description of the diffusion in dense liquids composed of non-spherical molecules.   
The success of this model also provides a powerful tool for studying water in confined geometries based on the concept that relates diffusion to the local structure.

Dynamics of dense liquids is known to be dominated by the mechanism called ``cage diffusion'' \cite{Cohen} where an elementary  motion of a diffusing particle can only be possible as a part of the rearrangement of the cage of its immediate neighbors. This mechanism was recognized in connection with the special feature in the neutron scattering spectra  (de Gennes narrowing   \cite{degennes}. In this way that the diffusion in liquids is coupled to the structural relaxation.\\

The process of structural relaxation in a liquid can be viewed as a random walk in the 3N-dimensional space of its configurations where $N$ is the number of particles. It is assumed that an elementary step in this process is a local particle rearrangement moving the system to a new configuration point. It only happens if the destination configuration is open the probability of which can be expressed through the excess entropy as $e^{S_{ex}}$. Therefore, the diffusion coefficient is supposed to scale universally as  \cite{dzugutov1,dzugutov2}

\begin{equation} \label{scaling_law}
 D = D_0 e^{s_{ex}}
\end{equation}

The basic postulates of this conjecture follows from the Arrhenius law for liquid diffusion:

\begin{equation} \label{arrhenius}
 D = D_0 e^{\Delta A/T}
\end{equation}

where $\Delta A$ is the free-energy barrier for an elementary diffusive step and $T$ is temperature.  $\Delta A =  \Delta u - T  \Delta s$ where $\Delta u$ is the enthalpic part of the free-energy barrier, and $ \Delta s$ is the respective entropic barrier. The first postulate is that liquid diffusion  can be universally  presented in terms of the diffusion of the liquid of Hard spheres (HS). In this case the notion of energy is missing and $\Delta A =  \Delta s$. Next, we assume that the entropy barrier is equal to the excess entropy (per particle) $s_{ex} = \log (  \Omega / \Omega_{PG})$ where $\Omega$ is the available volume of the system configuration space and $ \Omega_{PG}$ is the respective perfect gas volume. This assumption means that all the available configuration space sites are accessible for the diffusion. In this way we arrive to the scaling law, Eq. 1.

The excess entropy can be expanded in terms of the contributions from  n-body correlation functions as $s_{ex} = s - s_{pg} = \sum_{n=2}s_n$, where it is possible to consider only the two-body approximation. For an isotropic liquid environment two-body correlation function can be reduced to  the spherically symmetric  radial distribution function $g(r)$, and the respective two-body excess entropy approximation  can be expressed as  \cite{evans, raveche, mountain, wallace}

\begin{equation} 
 s_2= 2 \pi \rho \int_0 ^\infty \{ g(r) \ln [g(r)] - [g(r)-1] \} r^2 dr
\end{equation}

Using this form of the excess entropy, and by expressing the time scale in terms of the HS collision frequency, the scaling law in Eq. \ref{scaling_law} relates the liquid diffusion coefficient to the pair correlation function. In this way, it established a universal relation between liquid structure and diffusion  \cite{dzugutov1,dzugutov2}. This law  was found to be successful in describing  diffusion in a wide variety of simple liquids \cite{yokoyama98, babak,truskett_excess,jakse16} as well as in the solid ionic conductor $AgI$ \cite{dzugutov1}.  Also, a universal relation between thermodynamic entropy and Kolmogorov–Sinai entropy was found using this approach \cite{K-sinai} Here we shall use this scaling law to analyze the origin of the enhancement of water diffusion near a hydrophobic confining surface.

The change of local excess entropy $s_2$ due to the proximity of a confining surface can be accounted by the loss of structural correlations truncated by the surface. This is obtained by evaluating the missing part of $s_2$ which represents the correlations of the central particle with the particles that happened to be within the segment of the  sphere beyond the wall, and it results in the following expression\cite{agosta_entropic_2024}: 

\begin{equation} \label{Delta_s2}
\Delta s_2(d) = \pi \rho \int_0 ^\infty \{ g(r) \ln [g(r)] - [g(r)-1] \} (r-d) r dr
\end{equation}

We showed in our previous study\cite{agosta_entropic_2024} that it is possible to further correct this expression  taking into account the variation of $s_2$ due to the change of the local density and  g(r) when approaching the confining wall. We will call this modification $\Delta s_2^{mod}$
According to the scaling law (Eq. \ref{scaling_law}) this reduction of the absolute value  of $s_2$  due to the loss of correlation beyond the confining wall is expected to result in the enhancement of the diffusion rate for the particles near the wall which can be estimated as

\begin{equation} \label{D_s2}
  D(d)/D_b = e^{\Delta s_2(d)} 
\end{equation}

where $D(d)$ and $D_{b}$ are the diffusion rates at the distance $d$ from the wall and in the bulk liquid, respectively.

\begin{figure}[h!] 
\includegraphics[width=1.0\columnwidth] {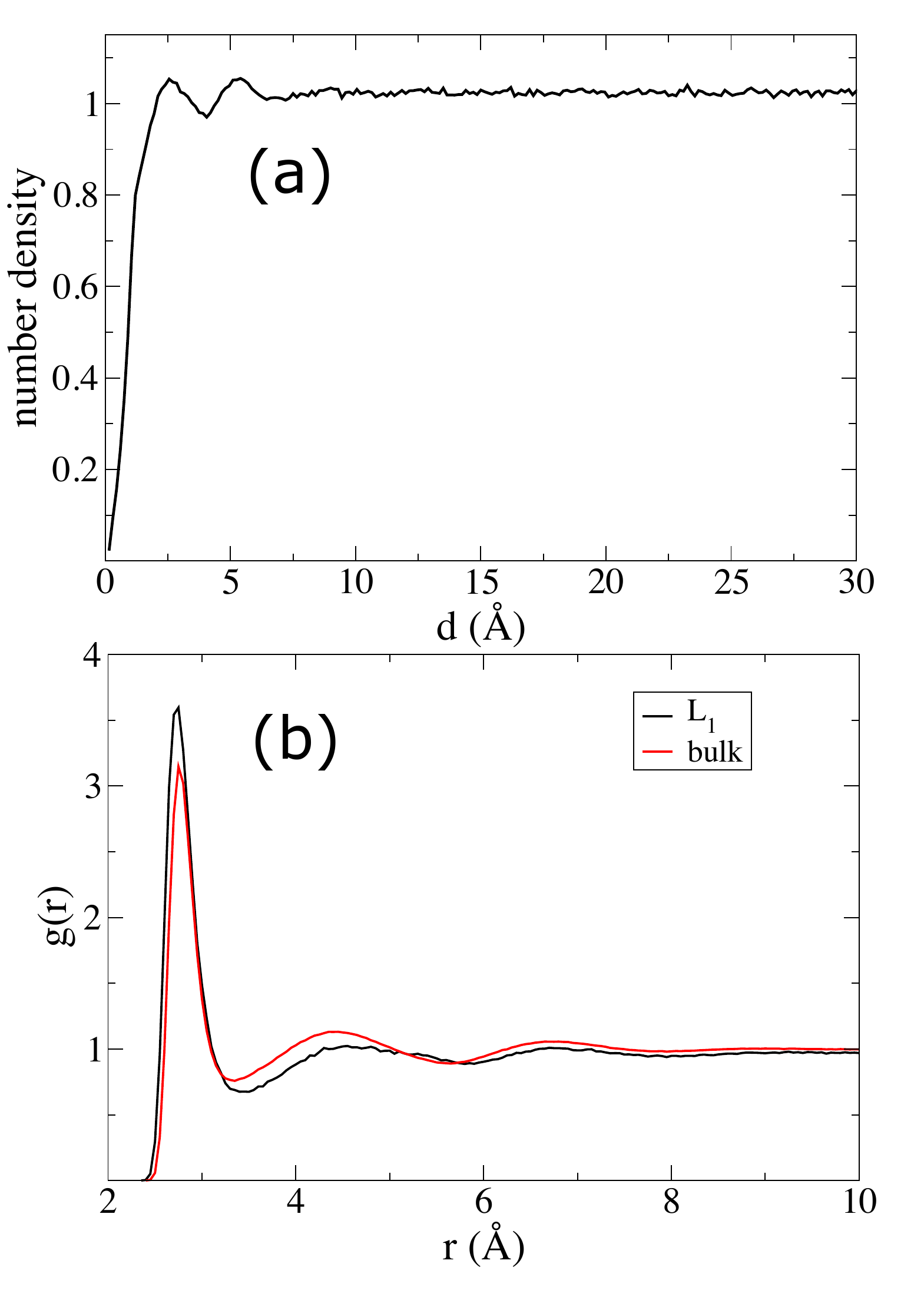}
\caption{ (a) The density profiles along the direction perpendicular to the wall for the confined liquid water and (b) the radial distribution function of the water layer centered at the first density peak and for the respective bulk.}
\label{plotdensity-rdf}
\end{figure}

In order to test the described model predicting the effect of the confining wall on the liquid diffusion we performed an MD simulation using a classical model of  water confined  between two smooth flat purely repulsive walls.
We consider layers of the simulated liquid parallel to the confining wall, with the layer width chosen be the molecule radius. In each layer we consider lateral diffusion in the plane perpendicular to Z axis. The molecular dynamics simulations considered a box of liquid water containing 12000 water molecules at 300K and 1 atm. The simulations were performed with GROMACS software\cite{GROMACS} for a TIP4P \cite{tip4p_ew} water model with a time integration of 0.5 fs. The bulk liquid system was obtained by using 3D periodic conditions in the NVT ensemble equilibrated for 5 ns. The confined system was achieved by imposing an hard potential at the bottom and the top of the simulation box along the Z direction. The wall was introduced with PLUMED\cite{Tribello} software as a potential energy between the water atoms as $U_z(d) = k(d-d_0)^b$, where $d$ is the distance along the Z direction between a water atom and the wall, $d_0=0.5$ is the position of the wall, $k=10^9$ is the force constant and $b=4$ relates to the stiffness of the wall. The system was equilibrated during 5 ns.

In Fig. \ref{plotdensity-rdf}(a) we report the density profile of the oxygen atoms as a function of the distance from the wall. We observe a density gradient from the wall to the first density peak at 2.5 \AA \,  (which is approximately the measure of one water molecule diameter). Beyond the first peak close the intensity of the density fluctuations are rather negligible, as also reported in previous studies \cite{chandler_wall_ideal}. Consistently the radial distribution function of water molecules belonging to the layer of half diameter apart from the wall experience local structure rearrangement with respect to the bulk water structure. Overall the presence of the wall seems to affect only the water molecule within one molecule diameter ($\approx$ 3 \AA) from the surface.

\begin{figure}[h!] 
\includegraphics[width=1.0\columnwidth] {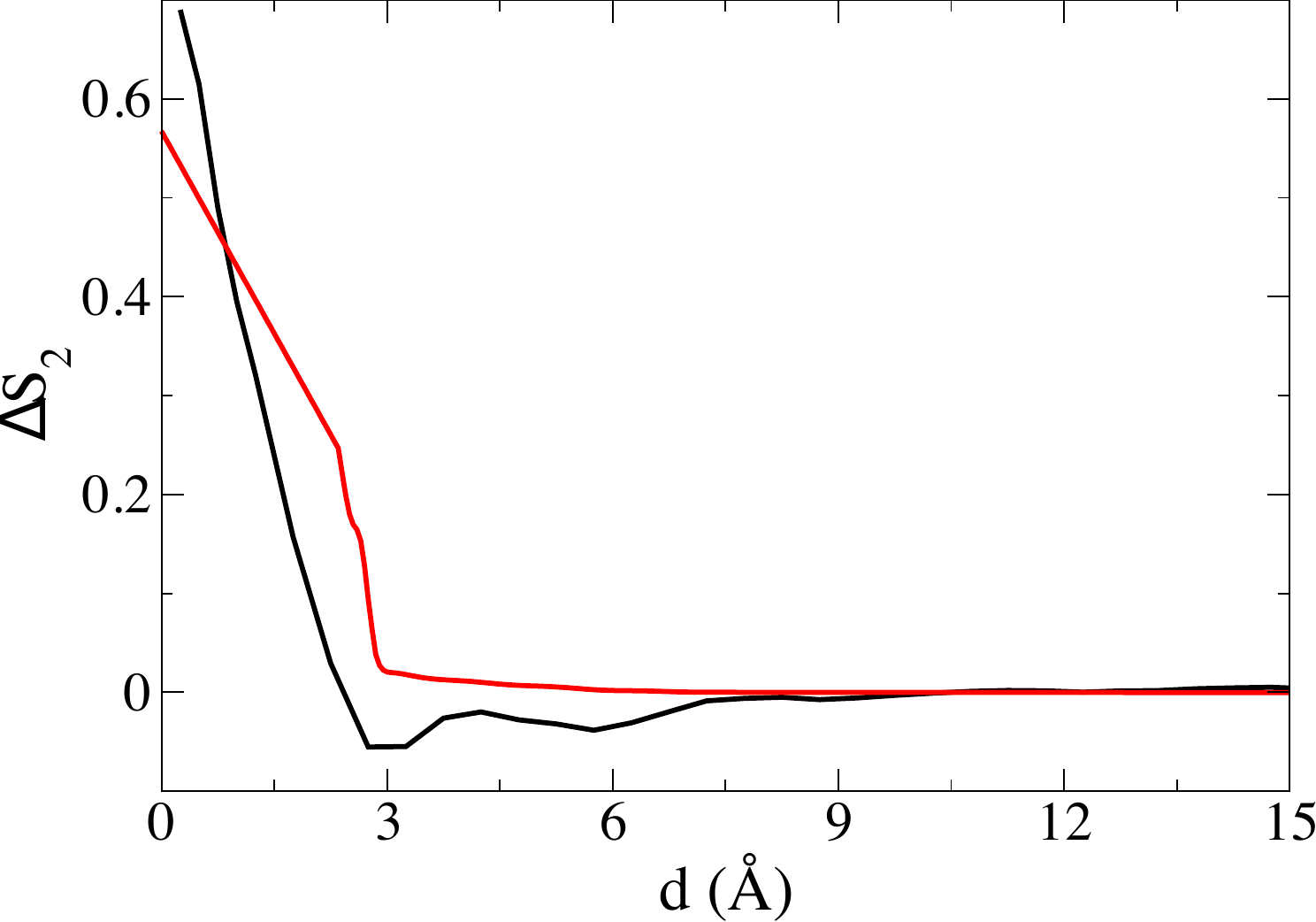}
\caption{Excess entropy variation as function of the distance $d$ from the confining wall (relative to the bulk value). The red curve shows the variation of $s_2(d)$ caused by the truncation of bulk g(r) imposed by the wall, as illustrated in  Eq. \ref{Delta_s2}. The black curve is the correction to $s_2(d)$ which includes the influence of the local density as shown is Fig. \ref{plotdensity-rdf}.}
\label{plot_S2}
\end{figure} 

We can now use the  theoretical model presented above to estimate how presence of a confining wall can increase the diffusion coefficient $D(d)$ in the liquid layer  at the distance $d$ from the wall as compared with that in the bulk liquid  $D_b$. In Fig. \ref{plot_S2} we show how this quantity depends on the distance from the wall. As one could presume from the purely geometric considerations,  the observed  effect of the proximity to wall upon $S_2$ is quite significant at the distance of the first layer, but it is quite short-ranged being limited to the first couple of layers.   

The change of the excess entropy as a function of the distance from the wall having been calculated, we can now estimate the expected enhancement of the diffusion rate according to Eq. \ref{D_s2}, and compare it with the actual variation of the  2D lateral diffusion rate within layers at different distances from the wall. The components of the diffusion coefficient parallel to the wall are evaluated from the mean-square displacement (MSD) corrected by the probability for a particle to stay within the layer during the time interval $t$ (ref).

\begin{equation} \label{msd} 
  D_{xy} (d) = \frac{1}{4} \lim_{t \to \infty}
  \frac{\text{MSD}_{xy} (d, t)}{t P(d, t)} \, ,
\end{equation}

where the MSD is defined in a layer at the distance $d$ from the wall. $P(d, t)$ is the probability that a particle stays in the layer within time interval $t$\cite{supercooled}. In all layers the latter time was found to be sufficient for the particles to exhibit long enough linear regime of MSD needed for the reliable calculation of the diffusion coefficient.

\begin{figure}[h!] 
\includegraphics[width=1.0\columnwidth] {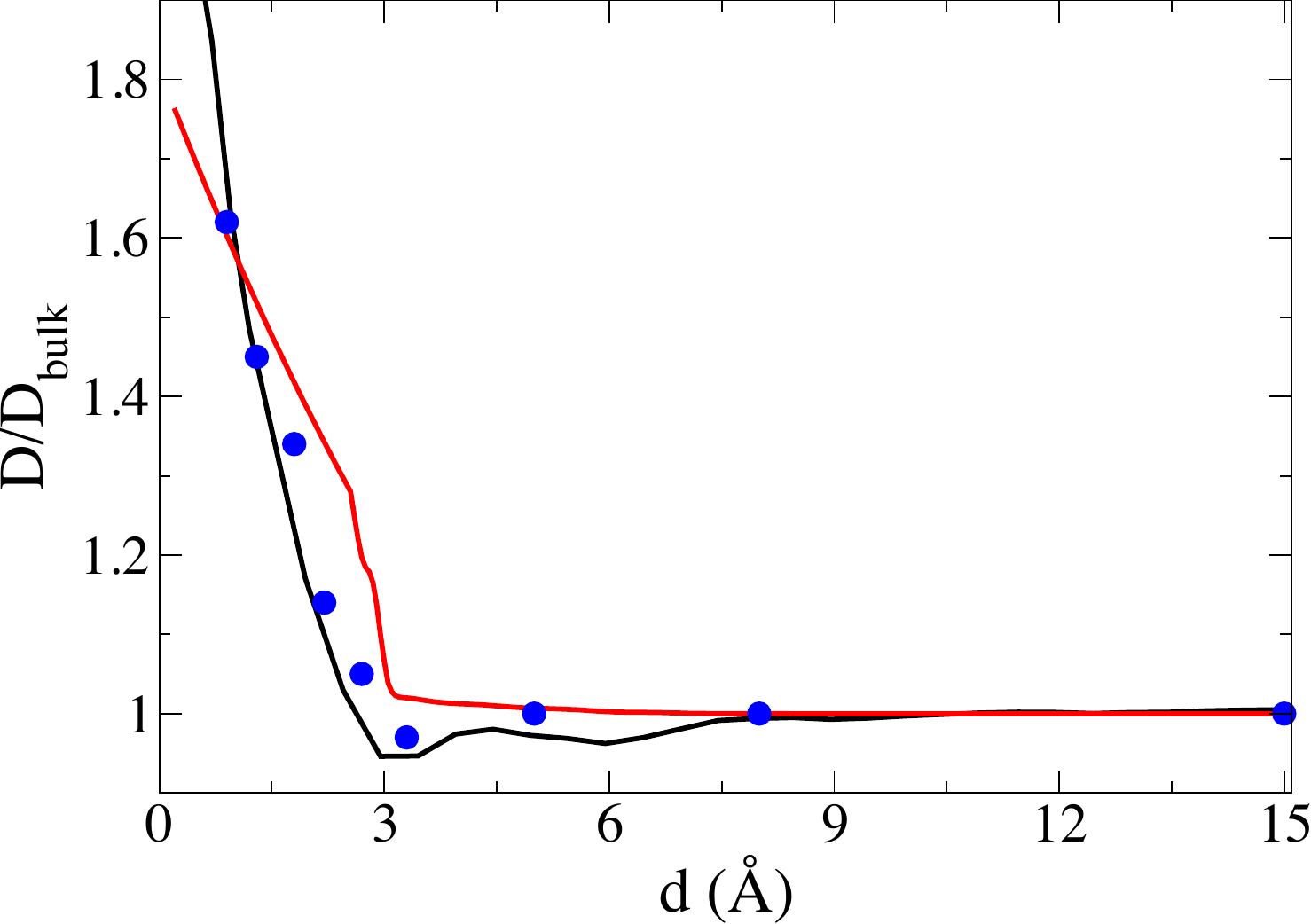}
\caption{Diffusion rate enhancement at the distance $d$ from the confining wall (relative to the bulk value). Dots, diffusion coefficient values in the layers of water calculated in molecular dynamics (MD) at the indicated distance from the confining wall. The dot positions are defined as the middle position of the respective layer thickness. The curves indicate the  diffusion coefficient as a function of the distance from the wall calculated  according to the entropy scaling law, Eq. \ref{D_s2} (red curve) and the relative correction to it due to local density fluctuations (black curve).}
\label{plot_D}
\end{figure}

In Fig. \ref{plot_D} we compare the lateral diffusion coefficient as a function of the distance from the wall with the bulk 2D diffusion. The effect of diffusion enhancement induced by the proximity of the wall is quite apparent.
Consistently, we observe the same increasing enhancement factor when we estimate the  diffusion rate from the theoretical prediction based on the excess entropy scaling law in the Eq. \ref{D_s2}. The observed effect is quite significant at the distance of the first water layer, but it decays to the bulk value beyond it, which is consistent with the short range structural changes described for Fig. \ref{plotdensity-rdf}.  

It is of interest to compare these results with the earlier reported results of the study of the confinement-induced  diffusion enhancement in the Lennard-Jones (LJ) liquid \cite{agosta_entropic_2024}. We can see that oscillations in the water radial distribution function and the density profile are shorter-ranged and less pronounced  than those observed in the LJ liquid. This can apparently be explained by the geometric frustration involved in the dense packing of anisotropic water molecules as compared with the more regular hard sphere-like  packing of the LJ particles. Accordingly, the confinement-induced  diffusion enhancement effect in water is of shorter range than the respective enhancement observed in the  LJ liquid.

Based on these results we propose a general picture of water behavior in the proximity of a confining surface. It is determined by the superposition of two independent effects, enthalpic and entropic. The enthalpic effect arises from the interaction forces between the water molecules and the surface, and, for obvious reason, it impedes the diffusion. This interaction effect apparently depends on the chemical nature and morphology of the surface. If it is strong, the surface is hydrophilic, a smooth surface with weak or non-existent interaction with water is classified as hydrophobic.  The diffusion enhancement effect that we investigated here is of purely geometric nature and therefore it is universal and independent of the surface properties.  Whether or not the entropic diffusion enhancement caused by the proximity of a confining surface  overwhelms its impediment caused by the interaction with the surface entirely depends on how strong the latter is.

Perhaps the most conceptually significant conclusion of the results we presented here concerns the universality of the scaling law relating liquid diffusion to the excess entropy. So far it has only been successfully tested on simple liquids \cite{babak,yokoyama98,truskett_excess}. Its tests on a water model were inconclusive \cite{herrero_excess, excess_gordon, agarwal_excess, mittal_excess,truskett_excess2,sahu17}. In these papers it was suggested that the entropy scaling should be modified as $\propto e^{\alpha s_2}$, $\alpha$ considered to be a fitting parameter. Such a modification is obviously inconsistent with the basic concept of the entropy scaling law \cite{dzugutov1,dzugutov2}. Indeed, $e^{s_2}$ quantifies the number of available configurations, whereas $e^{\alpha s_2}$ has no clear physical meaning if $ \alpha \neq 1$.  The observation that the water diffusion in the proximity of confining wall can be accounted for by the Eqs.  \ref{Delta_s2} and \ref{D_s2} makes it possible to conclude that the scaling law, in its original form, is valid for the description of water diffusion. The discrepancies observed in other tests \cite{herrero_excess, excess_gordon, agarwal_excess, mittal_excess,truskett_excess2,sahu17} can possibly be attributed to the estimation of the collision frequency involved in the prefactor \cite{babak,sahu17}. These results show that the universality of the scaling law which which has so far only been successfully tested on simple liquids can be extended to involve the liquids composed of non-spherical molecules with anisotropic interaction.

Other remarks are in order:\\

1. The scaling law for liquid diffusion used in this study describes the rate of the local relaxation dynamics. This means that the diffusion enhancement near the confining hydrophobic wall must also be accompanied by the reduction of viscosity \cite{abramson_excess}. According to the Stokes-Einstein law which relates the self-diffusion coefficient of a particle suspended in a viscous liquid to the liquid viscosity  this implies that the diffusion of nanoparticles with hydrophobic surface suspended in water must also be enhanced.

2. The theoretical model interpreting the diffusion enhancement in water near confining hydrophobic walls in terms of the excess entropy scaling law can be easily extended to explain similar diffusion enhancement near other kinds of non-interactive surfaces, including the surface separating the coexisting phases of  liquid water and water vapor \cite{diff_water_vap,water_dyn_hydrophobic}. 

3. It is quite clear from the geometrical arguments which were presented above that the diffusion enhancement effect in liquids induced by the proximity of a confining surface is expected to be stronger if the surface shape is concave with the curvature radius of nanometer scale. This conjecture is of significance for the applications like water dynamics in nanoporous materials \cite{water_slip_nanotubes,nanopore_dyn}.

4. As we have shown, the diffusion enhancement in water near a hydrophobic confining surface can be assessed, with a good accuracy, from the excess entropy calculated from the bulk radial distribution function $g(r)$. The latter is a Fourier transform of the structure factor $S(Q)$ which is readily available from the diffraction experiments. The practical significance is that it makes it possible to avoid the problems involved in direct measurements of water diffusion in quasi-2D layers near the surface.

5. The observation that the variation of the diffusion coefficient from bulk to the hydrophobic interface is proportional only to the change in entropy implies that the activation free energy for such diffusion mechanism, according the Arrhenius low in Eq. \ref{arrhenius}, has a zero enthalpic contribution. This is fundamental because it relates with the assessment of hydrophobic properties of surfaces according the Young-Dupre relation \cite{Young-Dupre}.

In summary, we report a molecular dynamics simulation of water near a smooth non-interactive confining wall which imitates a hydrophobic surface. Evidence is presented that a significant diffusion enhancement effect observed in the water layers near the confining wall can be successfully accounted for on quantitative level using the entropy scaling law for the liquid diffusion \cite{dzugutov1,dzugutov2}. We demonstrated that the observed diffusion enhancement is  a result of the reduction of (the absolute value of) the local excess entropy caused by the truncation of the structural correlation of a diffusing particle by the confinement. A conceptually significant conclusion of these results is that the scaling law we successfully tested here can be extended to the description of the diffusion of liquids composed of non-spherically symmetric particles with anisotropic interaction forces. It is also shown that the diffusion enhancement in  water in hydrophobic confinement can be assessed from the diffraction data on bulk water.

\bibliography{references}

\end{document}